\documentclass[%
 reprint,
 amssymb,
 prl,
 superscriptaddress,
 showpacs
]{revtex4-2}
\usepackage{graphicx}
\usepackage{dcolumn}
\usepackage{bm}
\usepackage{amsmath,latexsym}
\usepackage{csquotes}
\usepackage{siunitx}
\usepackage[english]{babel}
\DeclareMathOperator{\sinc}{sinc}
\usepackage{float}

\usepackage{braket}
\usepackage[normalem]{ulem}
\usepackage[caption=false, justification=centerlast]{subfig}
\usepackage{todonotes}

\usepackage[%
colorlinks=true,
urlcolor=blue,
linkcolor=blue,
citecolor=blue
]{hyperref}

\newcommand{\UOL}{Department of Physics, The University of Liverpool, Liverpool, L69 3BX, United Kingdom}
\newcommand{\CI}{Cockcroft Institute, Warrington WA4 4AD, United Kingdom}

\newcommand{\UOM}{Department of Physics and Astronomy, The University of Manchester, Manchester M13 9PL, United Kingdom}

\newcommand{\HIJ}{Helmholtz Institute Jena, Fröbelstieg 3, 07743 Jena, Germany}
\newcommand{\GSI}{GSI Helmholtzzentrum für Schwerionenforschung GmbH, Planckstraße 1, 64291 Darmstadt, Germany}

\newcommand{\ICMUV}{ICMUV, Universidad de Valencia, 46071 Valencia, Spain}

\newcommand{\FUHC}{Federal University of Health Sciences of Porto Alegre, Porto Alegre, RS, 90050-170, Brazil}

\newcommand{\PKU}{Center for Applied Physics and Technology, HEDPS, and SKLNPT, School of Physics, Peking University, Beijing 100871, China}

\begin{document}
	      

\title{Coherent synchrotron radiation by excitation of surface plasmon polariton on near-critical solid microtube surface}

\author{Bifeng Lei}%
\email{bifeng.lei@liverpool.ac.uk}
\affiliation{\UOL}
\affiliation{\CI}

\author{Hao Zhang}
\affiliation{\UOL}
\affiliation{\CI}

\author{Daniel Seipt}%
\affiliation{\HIJ}
\affiliation{\GSI}

\author{Alexandre Bonatto}
\affiliation{\FUHC}
\author{Bin Qiao}
\affiliation{\PKU}
\author{Javier Resta‐L\'{o}pez}
\affiliation{\ICMUV}
\author{Guoxing Xia}
\affiliation{\UOM}
\affiliation{\CI}

\author{Carsten Welsch}
\affiliation{\UOL}
\affiliation{\CI}

\date{\today}

\begin{abstract}
Coherent synchrotron radiation (CSR) is crucial for the development of powerful ultrashort light sources.
We present a mechanism for generating CSR in the form of generalised superradiance, based on surface plasmon polaritons (SPPs), which are resonantly excited on a solid, near-critical-density inner surface of a microtube.
A high-intensity, circularly polarised laser pulse, propagating along the microtube axis, efficiently couples the cylindrical SPP modes. This process creates azimuthally structured, rotating electromagnetic fields.
These rotating fields subsequently confine, modulate, and directly accelerate surface electrons to emit CSR in the Vavilov-Cherenkov angle.
We further demonstrate that by improving the azimuthal symmetry of these electrons, the helical modulation enables CSR emission across all azimuthal directions in the form of isolated harmonics, significantly enhancing radiation intensity even when full coherence is imperfect.
Our full 3D Particle-in-Cell simulations indicate this scheme can generate X-rays with coherence enhanced by up to two orders of magnitude compared to incoherent emission.
The challenges to experimentally realise this scheme are discussed, including the need for high-contrast lasers to prevent pre-plasma formation and the demanding tolerances for microtube fabrication and alignment, while these challenges are not beyond the scope of existing or near-future experimental capabilities.
\end{abstract}

\pacs{Valid PACS appear here}
\maketitle


Coherent synchrotron radiation (CSR) represents a unique regime of electromagnetic (EM) radiation emitted when relativistic charged particles—typically electrons—undergo accelerated motion in a collective, spatially ordered configuration~\cite{Nodvick:1954aa}.
Its exceptional properties, such as high brightness and broadband emission, offer significant advantages for applications requiring high-power radiation sources in the terahertz (THz) and far-infrared regimes, such as ultrafast spectroscopy~\cite{Nakanishi:1998aa}, particle acceleration diagnostics~\cite{Lou:2020aa, Kim:2021aa}, and advanced light sources~\cite{Couprie:2014aa, Shin:2021aa}.

The rapid progress in relativistic laser-plasma interactions has promoted the emergence of new compact radiation sources~\cite{Teubner:2009aa, Corde:2013aa, Lei:2018aa, Kumar:2025aa}.
Furthermore, an unexplored regime of superradiance has been recently discovered where coherent radiation can be obtained by specific modulation of the bunch to create the luminal or superluminal quasiparticle as the radiator~\cite{Vieira:2021aa}, which is distinct from the conventional superradiance mechanism~\cite{Dicke:1954aa} and free electron laser (FEL)~\cite{Pellegrini:2016aa}.
A quasiparticle is a collective excitation of a group of underlying particles that behaves as an effective single entity~\cite{sagdeev1969nonlinear}, which can travel at any velocity without violating causality.
Due to the superluminal motion, the wavefronts of the radiation emitted at different times can intersect to form an optical shock or caustic. At the crossing angle, the spherical waves are phase-aligned and interfere constructively.
This new theory unlocks novel coherent radiation in plasma-based accelerators~\cite{Malaca:2024aa}.

\begin{figure}[h!]
	\centering
	\includegraphics[width=0.4\textwidth]{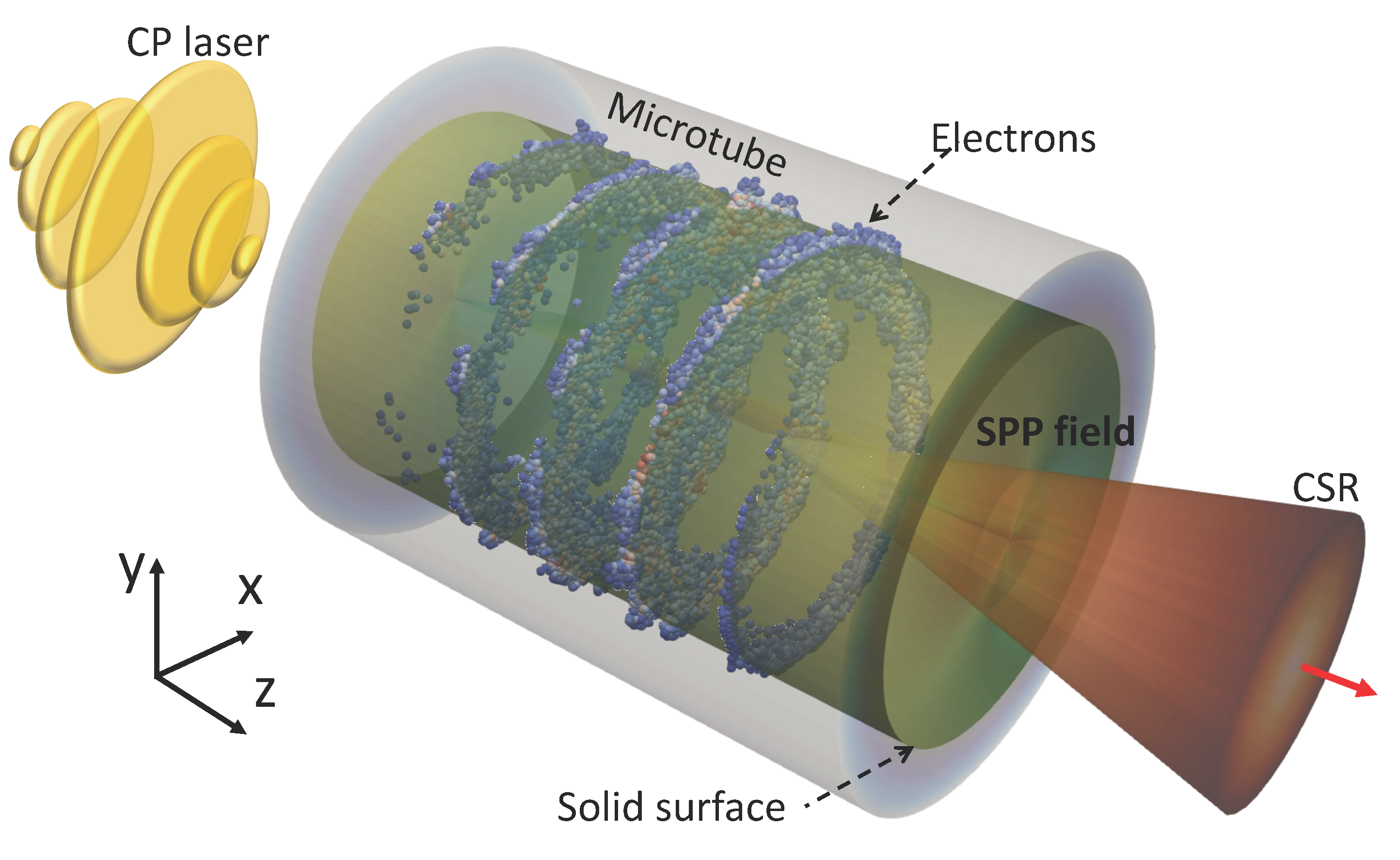}
	\caption{Schematic: a CP laser pulse (yellow) enters the vacuum channel of a microtube (grey) and excites SPPs while being scattered at the sharp vertical edge. The rotation mode $m=\pm 1$ (green-yellow) can efficiently couple with the laser field and accelerates the trapped electrons(sphere, blue-to-red colour presents the energy) into a spiral motion along the solid surface for the CSR emission (yellow-red cone).}
	\label{fig:illustration}
\end{figure}

A high-intensity laser pulse interacting with an interface between a metal and a dielectric medium (e.g., air or vacuum) can induce collective oscillations of free electrons. These oscillations, known as surface plasmons (SPs), are characterised by the plasma frequency, $\omega_p = \sqrt{4\pi n_e e^2 / m_e}$, where $n_e$ is electron density, $e$ is electron charge, $m_e$ is electron rest mass~\cite{Ritchie:1957aa}. 
The excitation of SPs has been recently proposed for generating TV/m-level wakefields in solid microtubes — a hollow structure composed of nanomaterials (e.g., a longitudinally aligned CNT forest) for relativistic particle acceleration~\cite{Lei:2025aa}.
When these SPs couple with incident EM waves, they form surface plasmon polaritons (SPPs)—hybrid light-electron modes that propagate along the interface, also termed surface plasmon waves~\cite{Cunningham1974eff}.
Direct excitation of SPPs on a smooth, infinite interface is inherently inefficient due to the momentum mismatch between the incident light and the SPP modes~\cite{Zhang:2012aa}. 
However, in finite-sized solid microtargets, sharp edges can serve as effective scatterers, supplying the necessary momentum to bridge this gap and enable efficient SPP excitation~\cite{Stegeman:1983aa}.   
The EM fields associated with SPPs can directly accelerate electrons along the target surface to relativistic energies.
Recent studies have shown that irradiating the edge of a flat microtape with a high-intensity, linearly polarised laser pulse can generate strong SPP, producing high-energy electrons and high-flux X-ray radiation~\cite{Shen:2024aa}. 

In this letter, we propose a comprehensive, self-contained and highly efficient plasma-based scheme to realise the principle of generalised superradiance, by using SPPs on a cylindrical surface to create a well-controlled field structure with a $\si{TV/m}$-level gradient which accelerates and modulates the self-trapped electron bunch acting as a superluminal quasiparticle.
We first show that SPPs can be directly excited on the cylindrical solid interface between plasma and vacuum by a circularly polarised (CP) laser pulse propagating paraxially through a microtube of near-critical density, as illustrated in Fig.~\ref{fig:illustration}.
The cylindrical geometry enables the CP laser to couple efficiently with a rotating SPP mode whose phase velocity approaches the speed of light in vacuum.
This resonant coupling provides $\si{TV/m}$-level SPP fields for the modulated trapping, confinement and stable acceleration of relativistic electrons over several Rayleigh lengths.
By matching the channel radius to the SPP field, the modulated, high-energy electrons rotate collectively as a quasiparticle near the surface, which produces CSR at the Vavilov-Cherenkov angle. 
We further demonstrate that, with an azimuthally symmetric distribution, the modulated beam can emit coherent radiation in all azimuthal directions and allow the harmonics to be well-separated.
Full 3D classical particle-in-cell (PIC) simulations have been used to verify the fundamental process. 
Numerical simulations demonstrate a significant coherence enhancement of photon emission compared to the incoherent case, with a radiation spectrum centred at a critical energy of several $\si{keV}$.
These results represent a significant step towards a viable experimental platform from the theoretical concept of generalised superradiance.

\begin{figure}
	\centering
	\includegraphics[width=0.48\textwidth]{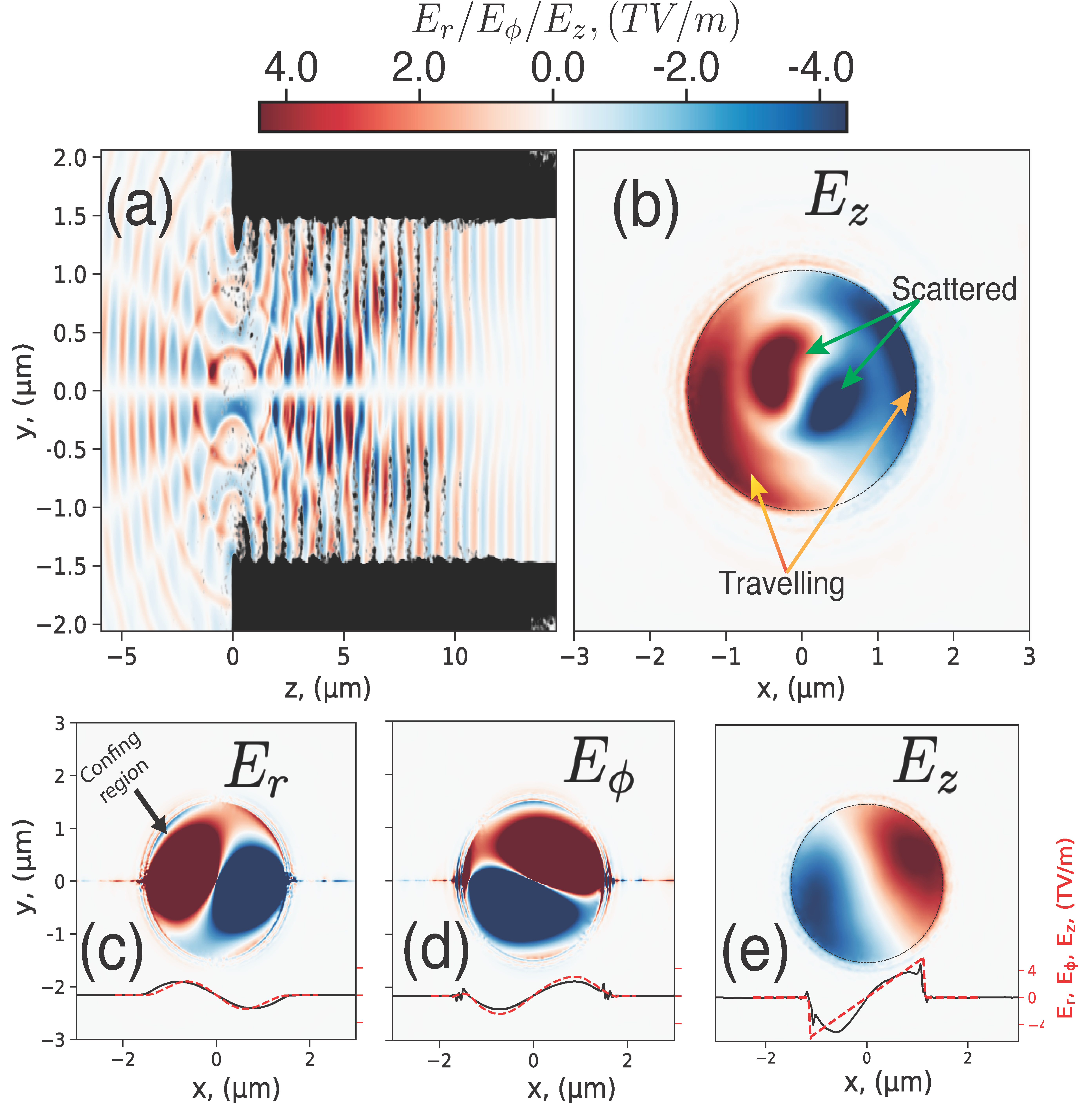}
	\caption{(a) Electron density distribution $n_e$ (black colour) and $E_z$ (blue-red colourmap) in $yz$-plane sliced at $x=0$ while the laser pulse entering the microtube. 
	(b) Snapshots of $E_z$ in $xy$-plane sliced at $z=5~\si{\mu m}$ corresponding to (a).
	(c)-(e) Snapshots of $E_r$, $E_{\phi}$ and $E_{z}$ in $xy$-plane at the same phase after the laser pulse propagating $z=41~\si{\mu m}$ in the tube.
	 The solid black and dashed red lines in each plot present the line plots of the corresponding field along $y=0~\si{\mu m}$ from PIC and analytical solutions, respectively.
	 }
	\label{fig:SPP_fields}
\end{figure}

A microtube can be made of nano-materials, e.g., a longitudinally aligned carbon nanotube (CNT) forest, which can provide the solid plasma with flexible density in the range of $10^{19} -10^{24}~\si{cm^{-3}}$~\cite{Yu:2009aa}.
The excited SPP and corresponding electric fields are shown in Fig.~\ref{fig:SPP_fields}, which are obtained from a PIC simulation where a microtube of radius $a=1.5~\si{\mu m}$ is used. 
A $20~\si{\mu m}$-long vacuum section is placed at the head of the target to initialise the laser pulse and provides a sharp annular edge at $z=0~\si{\mu m}$. 
As the bulk size of the tube is much larger than the inner radius of a CNT, e.g. up to several $\si{nm}$, the effect from the inner structure of CNTs is neglected.
A CP laser pulse of wavelength $\lambda_l=0.8~\si{\mu m}$ and normalised strength $a_0=6$ propagates along the axis in the $z$ direction and is focused to the entrance of the microtube, with root-mean-square (RMS) waist $w_0=2.0~\si{\mu m}$, and duration $\sigma_{\tau}=20.0~\si{fs}$. These parameters give a peak field of $24.0~\si{TV/m}$, or $13.6~\si{TV/m}$ on the inner surface. This is sufficient for the pulse front to instantaneously ionise electrons of the material within half a cycle.
Therefore, the material properties, e.g., lattice structure or optical absorption, are unimportant for electron dynamics. 
For femtosecond laser pulses, the interaction time is also in the order of femtoseconds. Significant hydrodynamic surface expansion, however, occurs on a much slower picosecond timescale. Therefore, for moderate laser strength with high contrast, the solid surface can be maintained during the interaction. 
These conditions allow the propagation of SPPs on the vacuum-plasma interface.

The simulations are carried out with the code WarpX~\cite{Fedeli2022}.
The dimensions of the moving window are $8~\si{\mu m} \times 8~\si{\mu m} \times 30~\si{\mu m}$, comprising $512\times 512 \times 1024$ cells in the $x$, $y$, and $z$ directions, respectively. Each cell contains 64 macro particles, which are sufficient to solve the electron dynamics effectively.
The microtube is initially modelled as a uniform distribution of neutral carbon atoms, with their ionisation energies adjusted to reflect CNT properties—specifically the weak $\pi$-bond and $\sigma$ (C–C) bond energies.
The initial carbon density is $n_{C0}=2\times 10^{21}~\si{cm^{-3}}$ which gives $\omega_p$ close to the laser frequency $\omega_l$.
The field ionisation is implemented using the Ammosov-Delone-Krainov (ADK) method ~\cite{Ammosov:1986aa}.

The vertical solid surface at $z=0~\si{\mu m}$ scatters the incident CP laser pulse and enables the direct matching with the SPP mode if the resonant condition $k_l=k_{spp}$ is satisfied, where $k_l$ and $k_{spp}$ are wave numbers of laser and SPP field, respectively. This condition is efficient near critical density, as shown in Fig.~\ref{fig:SPP_fields}(a) and (b).
The excited SPP fields can stably propagate with ultrahigh strength in $\si{TV/m}$-level, e.g. $4.5~\si{TV/m}$ in our simulations as shown in Fig.~\ref{fig:SPP_fields} (c)-(e).
Some electrons can be trapped into the SPP field and accelerated to relativistic speed by the longitudinal component $E_z$ and, at the same time, confined in a small vicinity near the surface by the transverse component $E_r$. 
The azimuthal component $E_{\phi}$ drives the modulated beam to rotate.

The PIC results can be confirmed theoretically by analysing the SPP excitation in cylindrical geometry, $(r, \phi, z)$~\cite{Schmeits:1988aa}. 
In general, a pure transverse magnetic (TM) mode (e.g. $B_{spp,z}=0$ and $E_{spp,z} \neq 0$) of SPP field can exist on the cylindrical surface.
The $z$-component of a SPP mode can be given by solving Maxwell equations, as $E_{spp,z}^{(m)}(r,\phi, z) = A f_m(\kappa r) e^{i(k_z z + m\phi - \omega t)}$, where $A$ is a constant representing the field amplitude. $m$ is the SPP mode number and $\kappa=\sqrt{k_z^2 - \epsilon_d \omega^2/c^2}$ with $k_z$ the wavevector along the tube axis, $\omega$ the frequency, $c$ is the speed of light in vacuum and $\epsilon_d$ permittivity of the medium.
$f_m(\kappa r)$ presents the radial field profile and depends on the medium properties.
In the vacuum channel, $r<a$, the field is presented by the modified Bessel functions of the first kind as $f_m(\kappa r)=I_m(\kappa_1 r)$ since the field is finite as $r\to 0$ and $\epsilon_d = \epsilon_1 = 1$.
In the plasma, $r>a$, the field is presented by the the modified Bessel functions of the second kind $f_m(\kappa r)=K_m(\kappa r)$ as the field decays to zero as $r\to \infty$ and $\epsilon_d =\epsilon_2= 1 - n_e / n_c(\omega)$ where $n_c$ is the critical density for the frequency $\omega$ as $n_c(\omega) = m_e \omega^2 / 4 \pi e^2$.
PIC simulation agrees well with the analytical solution as shown in Fig.~\ref{fig:SPP_fields}(c)-(e).
The other components can be obtained from the axial components using Maxwell's equations, for example, the electric components as	$E_{spp,r}^{(m)} = (ik_z/\kappa^2) \partial E_{spp,z}^{(m)} /  \partial r$ and $E_{spp,\phi}^{(m)} = i m k_z E_{spp,z}^{(m)} / \kappa^2 r$ for different regions, respectively.  The dispersion relation can be derived by considering the continuity of the SPP field at the interface of plasma and vacuum. It shows that in $m=1$ mode, $v_{ph}$ is very close to $c$ even for a low plasma density. 
This feature enables the $m=1$ mode to be used for relativistic electron acceleration in a near-critical plasma.
See more details in Appendix A and the Supplemental Material~\cite{Supplement_materials, Jackson:2021aa, Raynaud:2018aa}.

 \begin{figure}
	\centering
	\includegraphics[width=0.48\textwidth]{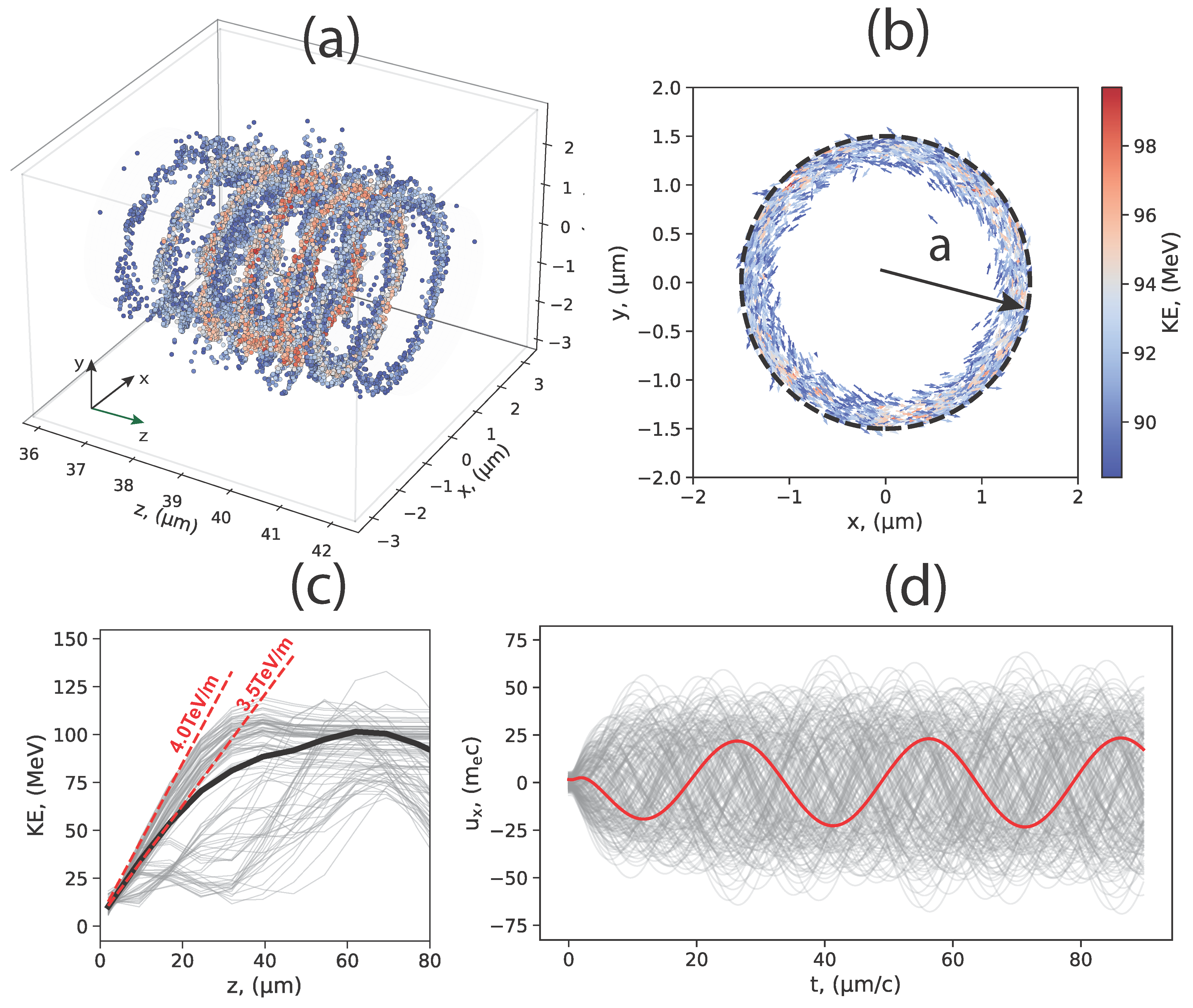}
	\caption{PIC results: 3D (a) and front (b) views of the high-energy ($\mathcal{E}>88~\si{MeV}$) electron beam modulation after the laser propagates $z=40~\si{\mu m}$ in the microtube. The arrows in (b) present the transverse momentum field ($u_x, u_y$).
	(c): Energy evolution of the high-energy electrons in the beam (grey lines) and the mean value (black line) as a function of propagation distance.
	(d) Trajectories of these electrons. The red line represents one of the electrons in the beam.
	}
	\label{fig:e_trajectory}
\end{figure}

The constant $A$ is proportional to the laser field amplitude $E_0$ and the overlap efficiency factor $\mathcal{O}(a, w_0, \kappa, m=1)$ as $A \approx \mathcal{O} \cdot E_0$.
From the PIC simulations as shown in Fig.~\ref{fig:SPP_fields}, $\mathcal{O}\simeq 0.19$, which implies that $100~\si{TW}$-level laser pulse can drive $\si{TV/m}$-level SPP field. The efficiency is orders of magnitude higher than that with a linearly polarised pulse, where $\mathcal{O} \sim 10^{-3}$~\cite{Lei:2025aa}.
The absolute value of $A$ depends on the incident laser power and the overlap integral between the incident laser field profile at the coupling region and the SPP mode profile, as 
\begin{equation}
	A \propto \int_0^a \int_0^{2\pi} \left( \mathbf{E}_{spp, \perp} \cdot \mathbf{E}_{laser, \perp}^*  \right)\bigg |_{z=0} r \, dr \, d\phi \mathrm{,} 
	\label{eq:coupling}
\end{equation}
where $\perp$ denotes components transverse to the z-axis.
Eq.~\eqref{eq:coupling} indicates that an efficient overlap requires $w_0$ to be comparable with the channel size $a$. This enables a larger laser focus to be matched with a wider tube.
It also indicates that, except $m=0$ mode, the SPP field with the azimuthal phase dependence $e^{im\phi}$ allows the enhanced coupling if the incident field can match the mode.
As a result, the efficient SPP excitation depends on how the laser field matches the azimuthal symmetry of the SPP mode.
 The coupling efficiency can be determined by $(\mathbf{E}_{spp}^{(m)} \propto e^{im\phi}) \cdot (\mathbf{E}_{\text{laser}}^* \propto e^{-il\phi}) \sim e^{im\phi} \times e^{-i l \phi} = e^{i (m-l)\phi}$ where $l$ is the azimuthal index of the laser field.
For a CP laser pulse, $l=\pm 1$ and the electric field $\mathbf{E}_{\text{laser}}(r, \phi, z, t) \approx E_0 (\hat{\mathbf{r}} + i \hat{\boldsymbol{\phi}}) g(r) e^{\pm i\phi} e^{i (k_l z - \omega_lt)}$
where $g(r)$ represents the spatial envelope.
The symbol $\pm$ denotes the right (RHC) or left (LHC) hand circular polarisation, respectively.
 The azimuthal structure of RHC or LHC polarisation can match the azimuthal phase dependence of the rotation SPP mode $m=1$ or $m=-1$, as $e^{\pm i\phi} \times  e^{\mp i\phi} = 1$, respectively.
 This leads to strongly enhanced coupling of these modes as seen in Fig.~\ref{fig:SPP_fields}.
 For the other modes, e.g. $m\neq 1$, integrating $e^{i (m\pm 1)\phi}$ over $2\pi$ gives zero, and the excitation is not efficient.
 This edge-coupling mechanism presented in this letter also enables the excitation of other modes by matching the azimuthal mode number $m$, for example, by a Laguerre-Gaussian laser~\cite{Jin:2023aa}.

The resonant excitation of SPP leads to highly efficient electron acceleration as shown in Fig.~\ref{fig:e_trajectory}(c), where the energy gain is $4.0~\si{TeV/m}$ in peak and the acceleration can be maintained for $40~\si{\mu m}$ or 3 Rayleigh lengths $Z_R$.
Due to the continuous trapping process, the electron beam is modulated to be helical, imprinted by the laser field. Therefore, the modulation frequency $\omega_m$ is $\omega_m=\omega_l$. The energy spectrum is broad.
Part of electrons, e.g. $18.7~\si{pC}$ or about $10\%$ of the total trapped electrons, can be accelerated to high energy $>90~\si{MeV}$ due to the early trapping.
These electrons can rotate around the channel axis while being confined closely along the interface, as shown in Fig.~\ref{fig:e_trajectory}(a) and (b).
The rotation is driven by the azimuthal components, and the frequency $\omega_r$ is estimated as $\omega_r = \sqrt{e(E_{\phi}+ \beta_z c B_r)/m_e a}$. With the value of fields from PIC simulations, $\omega_r\simeq 2 \pi\times 10^{13}~\si{Hz}$ or $\lambda_r=2\pi c/\omega_r \simeq 30~\si{\mu m}$, as shown in Fig.~\ref{fig:e_trajectory}(d).
The normalised phase velocity of the modulation is given as $\beta_{ph}\simeq 1 / (1-\omega_r/\omega_m) > 1$, which allows the constructive interference.

\begin{figure}
	\centering
	\includegraphics[width=0.48\textwidth]{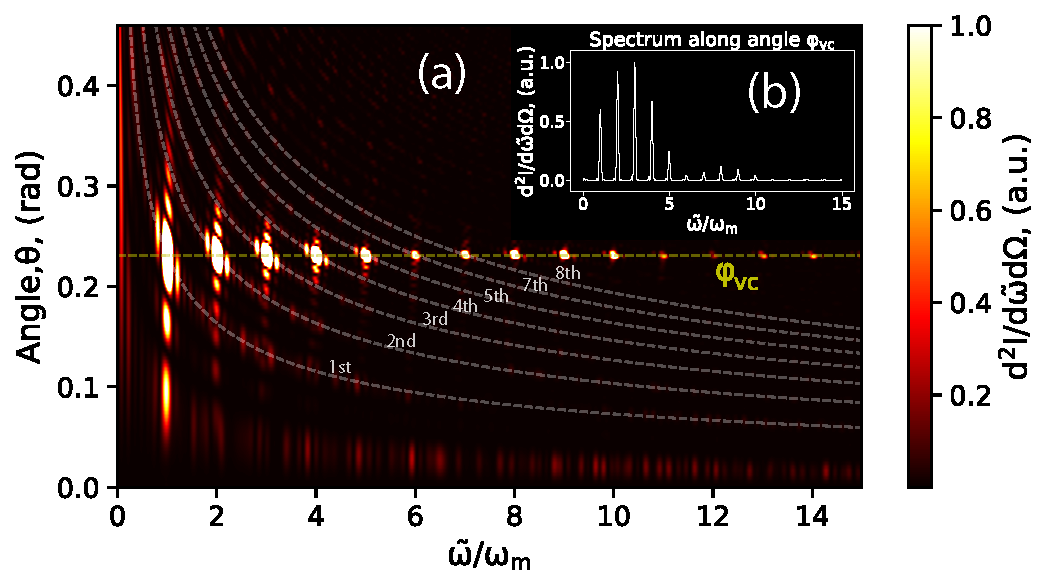}
	\caption{(a) Radiation spectrum coherently calculated by the trajectories of the electrons from a PIC simulation in Fig.~\ref{fig:e_trajectory}. Insert (b): Line plot along Vavilov-Cherenkov angle, $\varphi_{vc} \simeq 0.23$, indicated by horizontal yellow dashed line. The dashed white curves are theoretical resonant frequencies of different harmonics.}
	\label{fig:radiation_spectrum}
\end{figure}

The helically modulated oscillation of the electron beam can result in the CSR in the form of optical shock along the Vavilov-Cherenkov angle,  $\varphi_{vc}\simeq \arccos (1-\omega_r/\omega_m)$~\cite{Vieira:2021aa}. 
The radiation spectral lines are centred at frequencies $\tilde{\omega}_n=(n\pm 1) \omega_r/(1-\beta_z \cos \theta)$ where $n$ is an integer representing the harmonic number and $\beta_z$ the normalised longitudinal velocity of the beam. $\theta$ is the angle with $z$-axis in $yz$-plane. 
For $\beta_z\sim 1$, $\tilde{\omega}_n(n=0) \simeq \omega_m$ along in Vavilov-Cherenkov angle $\theta=\varphi_{vc}$. 
The radiation intensity is proportional to the square of the number of electrons inside the beam, $N$.
With the parameters used in this letter, $\varphi_{vc} \simeq 0.23$.
The radiation spectrum is shown in Fig.~\ref{fig:radiation_spectrum} (a) and (b), which is coherently calculated by using the wave optical method~\cite{Chubar:2006aa} with a virtual spherical detector in the far field where the radiation spikes appear in $\varphi_{vc}$ angle at the harmonic frequencies $\tilde{\omega}_n$.
The total energy radiated as CSR is $0.12~\si{mJ}$, which gives the energy conversion efficiency from the laser pulse to the CSR is $2.4 \times 10^{-4}$.
However, there are no advantages in modulating the bunch at the radiation frequency.

Now, we theoretically consider a specific modulation, an azimuthally symmetric modulation, where the initial phase of the electron trajectories is azimuthally uniform.
The trajectory of an electron (denoted by $j$) inside the electron beam can be written as $
	\bm{r}_j = (x_j, y_j, z_j) = a \left(\bm{\epsilon} e^{i \omega_r (t-t_{0j}) + i \psi_j} + c.c. \right)/2 + \bm{\eta} \beta_{z,j} c (t-t_{0j})$, where the polarization vectors are defined as $\bm{\epsilon}= (1, \pm i,0)$ and $\bm{\eta}=(0,0,1)$. $t$ is the retarded time and $t_{0j}$ the initial injection time.
In general, the initial phase $\psi_j$ depends on the injection time as $\psi_j=\omega_m t_{0j}$, which introduces the helical modulation with azimuthal symmetry. $\omega_m$ is the modulation frequency and $\omega_m=\omega_l$ in our case. 
Now, we further consider an additional condition, $\psi_j=\omega_m t_{0j}=2\pi j/N$, which implies that the electrons inside the beam are injected uniformly in the azimuthal phase. 
Note that this condition does not require the electrons to be close, as is the case with conventional superradiance~\cite{Dicke:1954aa}.
Since the beam is ultra-relativistic, $\beta_{z,j} \simeq \beta_z \simeq 1$.

\begin{figure}
	\centering
	\includegraphics[width=0.45\textwidth]{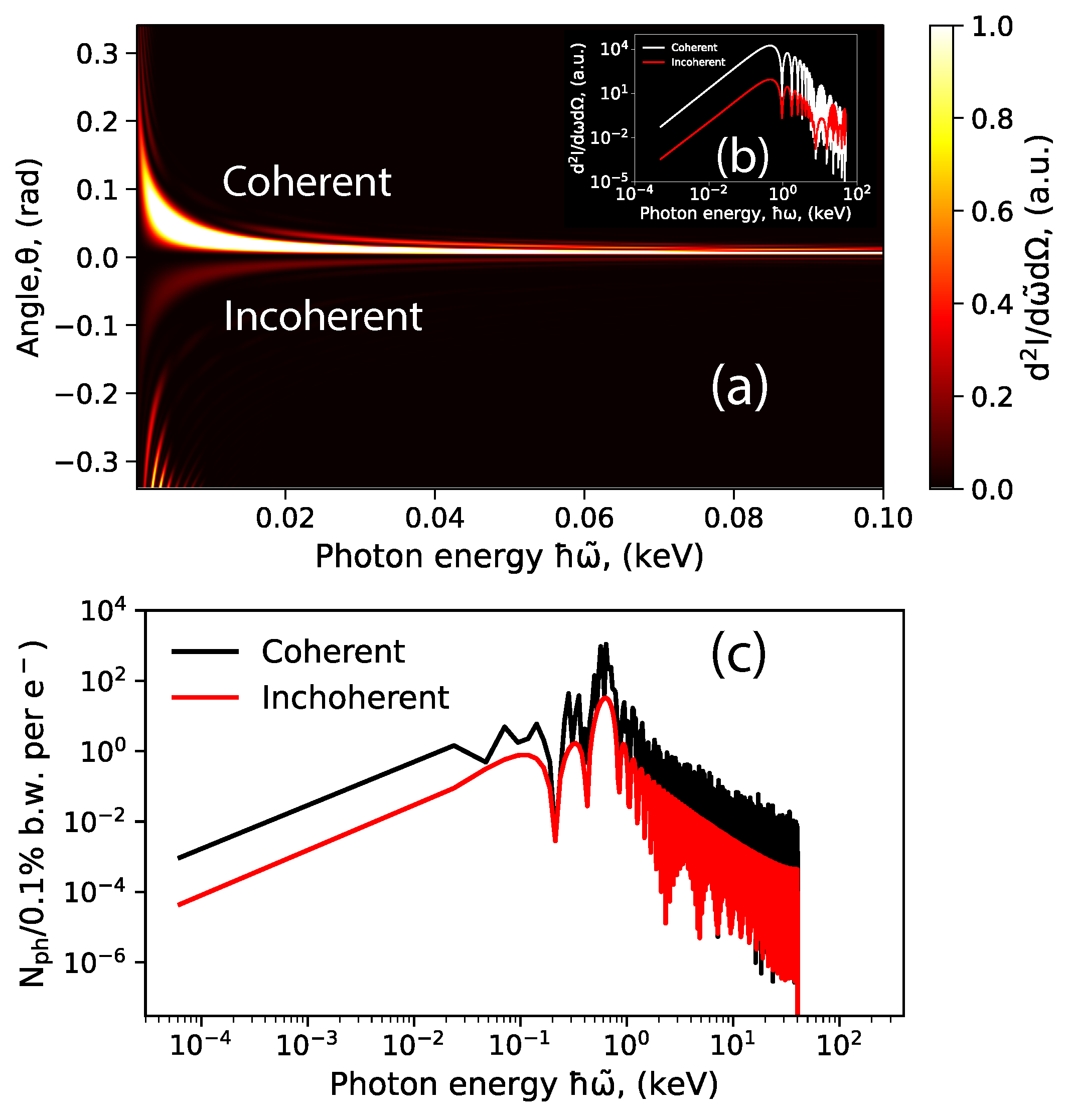}
	\caption{(a) Theoretical radiation spectrum calculated coherently (upper) and incoherently (bottom) by using an electron beam of $N=100$ electrons uniformly distributed over six modulation periods, or $4.8~\si{\mu m}$ long. The other beam parameters are the same as those from PIC simulations, as shown in Fig.~\ref{fig:e_trajectory}.
	Insert (b) Line plot of (a) along axis $\theta=0$ and $\phi=\pi/2$.
	(c) Line plot of onaxis ($\theta=0$ and $\phi=\pi/2$) spectrum in Fig.~\ref{fig:radiation_spectrum}(a).}
	\label{fig:azimuthal_rad}
\end{figure}

The radiation intensity $I$  per solid angle $d\Omega$ per frequency $d\tilde{\omega}$ in the far field can be written as $d^2 I/d \tilde{\omega} d \Omega = \mathcal{I}_0 \left |  \bm{\mathcal{A}} \right|^2$, where the constant $\mathcal{I}_0=e^2\omega^2 /4 \pi^2 c$.  
The radiation field $\bm{\mathcal{A}}$ is the contribution of the harmonics from the single electron and the modulation. It can be written as a superposition of the field from each single electron as $\bm{\mathcal{A}}  = \sum_{j=0}^{N} \bm{\mathcal{A}}_j$.
The radiation field from a single electron is given by $\bm{\mathcal{A}}_j  = \int_{-\frac{N_r \pi}{\omega_r}}^{\frac{N_r \pi}{\omega_r}}  \mathcal{M} \cdot \bm{\beta}_j(\bm{r}, t) e^{i \tilde{\omega} (t-\frac{\hat{n}\cdot \bm{r}_j}{c})} dt $
where $\hat{n}=(\cos \phi, \sin \phi \sin \theta, \sin \phi, \cos \theta)$ is the unit vector to the observer in the coordinate system and $N_r$ is the number of oscillation periods, $\phi$ is the angle with the $x$-axis in the $xy$-plane. $\mathcal{M}$ is a matrix defining the spatial distribution.
Using the modulated electron trajectories, the radiation field can be given by
\begin{equation}
\begin{split}
 \bm{\mathcal{A}} & =\frac{N N_r \pi}{\omega_r}   \mathcal{M} \cdot \sum_{n=-\infty}^{\infty}  i^{n} \bigg [ k_r a \bigg (\bm{\epsilon} J_{nN-1}(\zeta) \sinc(\omega_{-}) + \\
 &  \bm{\epsilon^*} J_{nN+1}(\zeta) \sinc(\omega_{+}) \bigg) +  2 \bm{\eta} \beta_z J_{nN}(\zeta) \sinc(\omega_z)  \bigg ]
	\label{eq:final_RE}
\end{split}
\end{equation}
with $\omega_{\pm} = N_r \pi (\omega_{nN}\pm \omega_r) /\omega_r$, $\omega_z = N_r \pi \tilde{\omega}_{n} /\omega_r$ and $\omega_{nN}= \omega(1-\beta_{z,j} \sin \phi \cos \theta)+nN\omega_r$. $\zeta=(\tilde{\omega} a/c) (\sin \theta \sin \phi -  \sin \theta \cos \phi)$.
It is seen from Eq.~\eqref{eq:final_RE} that the radiation spectral lines are centred at frequencies $\tilde{\omega}_n=(nN\pm 1) \omega_r/(1-\beta_z \cos \theta \sin \phi)$, which presents the CSR in all azimuthal directions rather than being confined to the angle $\varphi_{vc}$.
It also indicates that the harmonic numbers are given by $nN$. This implies the harmonics can be well separated. For sufficiently large $N$, such as a $\si{pC}$-level beam, only the fundamental harmonic ($n=0$) is present in the spectrum, as shown in Fig.~\ref{fig:azimuthal_rad}(a). This is physical, as multi-wave interference depends on the number of waves. See more details in Appendix B and Supplemental Material~\cite{Supplement_materials}. 
This characteristic offers a significant advantage in that the precise control over the radiation spectrum can be significantly enhanced.
The on-axis radiation critical frequency of an electron is $\tilde{\omega}_c \simeq 3 K_0 \gamma^2 \omega_r/2$~\cite{Kincaid:1977aa}. For an average $\gamma=100$ and oscillation strength $K_0=10$ as seen in Fig.~\ref{fig:e_trajectory}, this yields $\hbar \tilde{\omega}_c \simeq 6.2~\si{keV}$, which approximately aligns with the numerical simulations presented in Fig.~\ref{fig:azimuthal_rad} (c).
In realistic beams, such as those obtained from PIC simulations, the modulated electrons are distributed over a finite radial extent of $\Delta r \simeq 0.15~\si{\mu m}$, as shown in Fig.~\ref{fig:e_trajectory}(b). 
This spatial spread disrupts the azimuthal symmetry, thereby impeding the emission of fully coherent radiation.
A simple estimation of a Gaussian beam can give the form factor as $|F(\omega)|^2 \sim e^{-\Delta r^2/a^2}$, where $\Delta r$ reduces the amplitude of $|F(\omega)|$.
Nonetheless, partial coherence can still be achieved. This occurs when a subset of electrons satisfies the necessary symmetry requirement of $\psi_j$, leading to an enhanced radiation intensity, as shown in Fig.~\ref{fig:azimuthal_rad}(c), where the photon emission per electron is increased by almost two orders of magnitude compared to that predicted by incoherent calculations.
In summary, in comparison to the laser plasma wakefield acceleration (LWFA)-based sources, e.g. betatron radiation, our scheme offers distinct advantages in terms of coherence, resulting in a significantly higher photon flux per electron and brightness.
Although the total photon flux is below that of a free-electron laser (FEL) due to the limited charge accelerated in a single microstructure, it could offer several advantages, including micron-scale compactness, a broadband harmonic spectrum and robustness against beam parameters such as energy spread.
The unique combination of these properties makes it a compelling alternative to FELs for applications where they remain impractical, such as multi-spectral coherent pump-probe experiments at the laboratory scale or attosecond-scale diagnostics requiring a broad coherent bandwidth.

The experimental feasibility of this scheme depends on several practical efforts in target fabrication, laser operation and plasma diagnostics to optimise the SPP excitation and the helical modulation of the accelerated beam. 
While the main challenges include actively stabilised laser-target misalignment, polarisation quality, laser contrast and jitter, and target survivability, the technologies to address these issues are readily available in the state-of-the-art laboratories. 
The structured microtube target can be fabricated using advanced nano techniques~\cite{Seah:2011aa}, though the surface damage will limit the high repetition rate that needs to be further investigated.
Modern laser technologies can provide ultrashort optical pulses with high quality~\cite{Strickland:2019aa}.
The new generation of high-resolution laboratory diagnostics can provide deep insight into physical processes of SPP excitation, electron acceleration, and radiation generation~\cite{Downer:2018aa}.
With these technique advantages, our work introduces a novel approach to generating ultra-bright synchrotron radiation.

\begin{acknowledgements}
Javier Resta-López acknowledges support by the Generalitat Valenciana under grant agreement CIDEGENT/2019/058.
This work made use of the facilities of the N8 Centre of Excellence in Computationally Intensive Research (N8 CIR) provided and funded by the N8 research partnership and EPSRC (Grant No. EP/T022167/1). The Centre is coordinated by the Universities of Durham, Manchester and York.
\end{acknowledgements}

\appendix

\section{\textbf{APPENDIX A}: Dispersion relation of SPP in cylindrical surface} \label{appsec:AppendixA}
The dispersion relation can be derived by considering the continuity of the $E_z$ and $B_{\phi}$ of the SPP field at the interface of plasma and vacuum as
 \begin{equation}
 	\frac{I_m'(\kappa_1 a)}{\kappa_1 I_m(\kappa_1 a)} = \frac{K_m'(\kappa_2 a)}{\kappa_2 K_m(\kappa_2 a)} \mathrm{,}
 	\label{eq:disp_relation}
 \end{equation}
 where $\kappa_1=\sqrt{k_z^2 - \omega^2/c^2}$ and $\kappa_2=\sqrt{k_z^2 - \epsilon_2 \omega^2/c^2}$.
 For a large $a$, the dispersion relation in Eq.~\eqref{eq:disp_relation} resembles for $m=0$ as  $k=(\omega/c) \sqrt{\epsilon_2/(1+\epsilon_2)}$, which is the dispersion relation of SPP on the flat surface~\cite{Ritchie:1957aa}.
In the $m=0$ mode, the phase velocity $v_{ph}$ evolves similarly with that in planar geometry and only approaches $c$ as $n_e$ is much higher than $n_c$. 

\appendix
\section{\textbf{APPENDIX B}: Defination of matrix $\mathcal{M}$}
The matrix $\mathcal{M}$ is defined as the spatial distribution of the radiation field projected to the spherical detector. It is obtained from the formula
\begin{equation}
	\hat{n} \times (\hat{n} \times \bm{\beta}_j) e^{i \omega (t-\frac{\hat{n}\cdot \bm{r}_j}{c})} = \mathcal{M} \cdot \bm{\beta}_j e^{i \omega (t-\frac{\hat{n}\cdot \bm{r}_j}{c})}\mathrm{,}
\end{equation}
It can be written as
\begin{widetext}
\begin{align*}
\mathcal{M}  & = 
\begin{pmatrix}
	-(\hat{n}_y^2+\hat{n}_z^2)  & \hat{n}_x \hat{n}_y & \hat{n}_x \hat{n}_z \\
     \hat{n}_x \hat{n}_y  & -(\hat{n}_x^2+\hat{n}_z^2) & \hat{n}_y \hat{n}_z  \\
	\hat{n}_x \hat{n}_z  & \hat{n}_y \hat{n}_z & -(\hat{n}_x^2 + \hat{n}_y^2) 
\end{pmatrix} \\
& =
\begin{pmatrix}
-(\sin^2 \theta \sin^2 \phi + \cos^2 \theta)  & \sin^2 \theta \cos \phi \sin\phi & \sin \theta \cos \theta \cos \phi \\
\sin^2 \theta \cos \phi \sin\phi & -(\sin^2 \theta \cos^2 \phi + \cos^2 \theta) & \sin \theta \cos \theta \sin \phi \\
\sin \theta \cos \theta \cos \phi & \sin \theta \cos \theta \sin \phi & - \sin^2 \theta
\end{pmatrix} \mathrm{.}
\end{align*}
\label{eq:M_matrix}
\end{widetext}

\bibliography{cntrad.bib}

\end{document}